\documentclass[preprint,10pt,aip,apl,twocolumn,superscriptaddress]{revtex4}
\usepackage{graphicx}
\usepackage{amsmath,amssymb}
\usepackage{color}
\usepackage{gensymb}

\begin{document}

\title{Experiments on Maxwell's Fish-eye dynamics in elastic plates}

\author{Gautier Lefebvre}
\author{Marc Dubois}
\author{Romain Beauvais}
\affiliation{
Institut Langevin, ESPCI ParisTech CNRS UMR7587, 1 rue Jussieu, 75238 Paris cedex 05, France\\
}
\author{Younes Achaoui}
\affiliation{
Institut Fresnel, CNRS UMR7249, Aix-Marseille Universit\'e, Centrale Marseille, 
Facult\'e des Sciences de Saint-J\'er$\hat{o}$me,
Avenue Escadrille Normandie-Ni\'emen, 13397 Marseille Cedex 20, France
\\
}
\author{Ros Kiri Ing}
\affiliation{
Institut Langevin, ESPCI ParisTech CNRS UMR7587, 1 rue Jussieu, 75238 Paris cedex 05, France\\
}
\author{S\'ebastien Guenneau}
\affiliation{
Institut Fresnel, CNRS UMR7249, Aix-Marseille Universit\'e, Centrale Marseille, 
Facult\'e des Sciences de Saint-J\'er$\hat{o}$me,
Avenue Escadrille Normandie-Ni\'emen, 13397 Marseille Cedex 20, France
\\
}
\author{Patrick Sebbah}
\email[Contact: ]{patrick.sebbah@espci.fr}
\affiliation{
Institut Langevin, ESPCI ParisTech CNRS UMR7587, 1 rue Jussieu, 75238 Paris cedex 05, France\\
}

\date{\today}

\begin{abstract}
We experimentally demonstrate that a Duraluminium thin plate with a thickness
profile varying radially in a piecewise constant fashion as $h(r)=h(0)(1+(r/R_{max})^2)^2$, with $h(0)=0.5$~mm, $h(R_{max})=2$~mm and $R_{max}=10$~cm behaves in many ways as Maxwell's fish-eye lens in optics, since its imaging properties for a Gaussian pulse with central frequencies 30~kHz and 60~kHz are very similar to those predicted by ray trajectories (great circles) on a virtual sphere (rays emanating from the North pole meet at the South pole). However, refocusing time depends on the carrier frequency as a direct consequence of the dispersive nature of flexural waves in thin plates. Importantly, experimental results are in good agreement with Finite-Difference-Time-Domain simulations.
\end{abstract}
\maketitle

There is currently a keen interest in platonic metamaterials, which are structured thin plates within which flexural waves can follow some curved
trajectories, say around an obstacle \cite{farhat2009,stenger2012}, in ways
similar to what was achieved with electromagnetic invisibility cloaks
\cite{ulf2006,john2006}. In this vein, a flat lens was experimentally demonstrated around $10$~kHz using anomalous dispersion in a duraluminium
plate with a square array of circular air holes \cite{marc2013}. This work
is reminiscent of negative refraction observed for Lamb waves in a silicon plate with air inclusions \cite{Pierre2010}.
From a theoretical standpoint, one way to interpret a flexural flat lens
using tools of transformation platonics \cite{seb2010} is by mapping the
image plane onto the source plane. Such a space folding
\cite{Philbin2006} leads to Pendry's perfect lens \cite{Pendry2000}.

There are nonetheless other exciting transformation based lenses that do not require multivalued maps.
The fish-eye lens, which first appeared in a problem and its subsequent
solution published by James Clerk Maxwell in two issues of the Cambridge Dublin Mathematical Journal in 1853 and 1854 \cite{ulf1853}, is a sphere of radius
$R_{max}$ for which the refractive index varies according to
\begin{equation}\label{eq_index}
	n(r)=\frac{2}{1+(r/R_{max})^2},
\end{equation}
where $r$ is the distance from the center of the sphere. Inside
the lens, ray paths are circles and all rays from an object at $r_0$
converge at the image point $r_I = −r_0R_{max}^2/|r_0|^2$. This
heterogeneous lens also works in two-dimensions, its refractive
index still satisfies Eq.~\ref{eq_index}, and it has recently
attracted attention of researchers working in photonics \cite{ulf2009,blaikie,seb2010,ulf2011,merlin2011}, partly because
a fish-eye lens within a cavity is reminiscent of a time-reversed mirror \cite{lerosey}. Besides from that, it can be thought of as a virtual sphere stereographically
projected on a plane \cite{ulf1964}, what makes a transformation optics device.
In this letter, we experimentally explore the dynamics of an elastic plate whose thickness variation achieves the index profile defined by Eq.~\ref{eq_index}. It has indeed been suggested that plates of varying thicknesses make focussing effects of Lamb waves possible \cite{dehesa2014} and this provides a nice playground for transformation platonics.

\begin{figure}[]
\begin{center}
\includegraphics[]{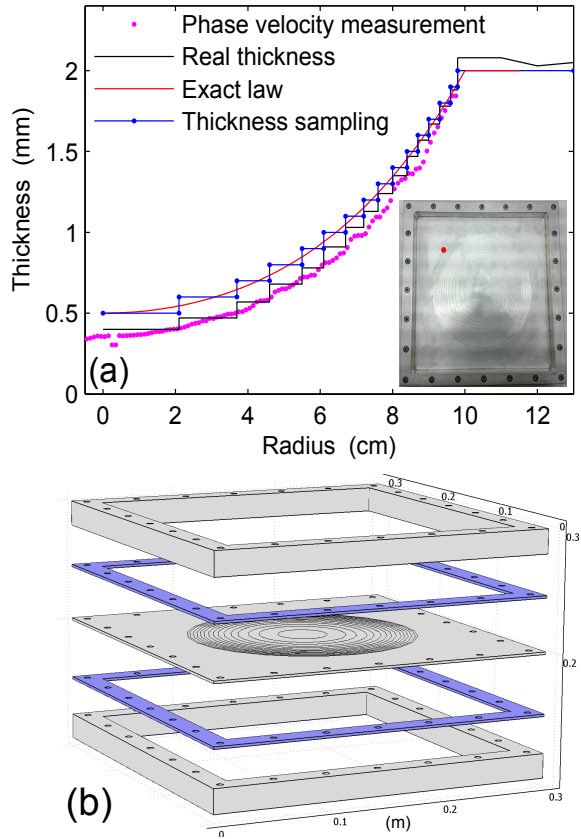}
\end{center}
\mbox{}\vspace{-1cm}
\caption{ (a): Thickness profile of the plate as a function of the distance from the center of the plate. The red curve is the theoretical thickness profile given by Eq.~\ref{eq_thickness}. The blue curve is the discretized profile used to design the plate. The black curve is the actual thickness profile obtained after machining and measured with a dial indicator.  The magenta dots represent the thickness measured with a method based on the local phase velocity \cite{etaix2010}. A picture of the whole setup is displayed in inset. The red dot marks the position of the source. (b): Sketch of the experimental setup. The plate with its fifteen equi-height circular rings is sandwiched between two metallic frames. The blue thin frames are layers of adhesive rubber introduced between the plate and the metallic frame. Various parts are screwed together.}
\label{fig_design}
\end{figure}

For harmonic excitation at pulsation $\omega$ in the low frequency regime, flexural waves in thin plates can be described by the Kirchhoff-Love Eq.~\ref{eq_kirchhoff} (i.e. when their wavelengths are much larger
than plates's thicknesses):
\begin{equation}\label{eq_kirchhoff}
	D \Delta^2 W - \rho h \omega^2 W = 0,
\end{equation}
where $W$ is the vertical displacement, $h$ the plate thickness and $\rho$ the mass density. 
From this equation, one obtains a quadratic dispersion relation valid for flexural waves in thin plates at low frequency:
\begin{equation}\label{eq_dispersion}
	\omega^2 = \frac{D }{\rho h} k^4 ,
\end{equation}
where $k$ is the wavevector modulus and $D=E h^3/12(1-\nu^2)$ is the flexural rigidity, with $E$ the Young's modulus and $\nu$, the Poisson's ratio. The refractive index is defined as the inverse of the phase velocity:
\begin{equation}\label{eq_phase_velocity}
	V_{\phi} = \frac{\omega}{k} = \sqrt[4]{\frac{E h^2 \omega^2}{12 \rho (1-\nu^2)}} .
\end{equation}

The physical parameters offer  several degrees of freedom to engineer at will the dispersion relation and the refractive index in thin plates \cite{Dehesa2013}, providing their variation is adiabatic for Eq.~\ref{eq_phase_velocity} to remain valid. For more details on this assumption, see \cite{SuppMat}. Here we choose to vary the thickness as it is technically a more workable parameter than $\rho$ or $E$. The refractive index is inversely proportional to the square root of the plate thickness, as seen from Eq.~\ref{eq_phase_velocity}. The index variation of Eq.~\ref{eq_index} required to design the Maxwell fish-eye lens translates here into the following circular thickness profile:
\begin{equation}\label{eq_thickness}
	h(r)=h(0)(1+(r/R_{max})^2)^2.
\end{equation}

When $R_{max}$ is set to 10~cm, plate thickness varies from $h(0)=0.5~$mm to $h(R_{max})=2~$mm. This continuous thickness variation is discretized into equi-height concentric rings as shown in Fig.~\ref{fig_design}b to simplify the machining. These rings were carved in a Duraluminium square plate (30~cm$\times$30~cm), using a MAZAK, VTC200B with 5 micron accuracy (see insert in Fig.~\ref{fig_design}b). Although small, the error is cumulative. The actual plate profile obtained after processing slightly departs from the theoretical profile (Fig.~\ref{fig_design}b), but the original profile is essentially preserved. We have also performed a measurement based on the local phase velocity \cite{etaix2010}, which reveals a very smooth profile. This suggests that the waves are not sensitive to the steep thickness variations, but only to the average profile. The sample arrangement is shown in Fig.~\ref{fig_design}a. The plate is held in between two Duraluminium frames, with inner dimensions of 26~cm$\times$26~cm. Layers of adhesive rubber (3.5 mm-thick) are inserted between the frame and the plate before screwing, in order to absorb outgoing waves and reduce reflections at the boundaries. An attenuation of 12~dB is achieved at $f$=30~kHz.

The acoustic emitter is a 1~cm-diameter piezoelectric ceramic disk (PKS1-4A1 MuRata Shock Sensor) bound at the edge of the largest ring (see inset of Fig.~\ref{fig_design}b), using Phenyl salicylate (Salol-melting point of $43~\degree~C$) to ensure acoustic coupling. The input signal is a digital-to-analog converted and amplified acoustic Gaussian wavepacket centered at $f_0$=60~kHz with a bandwidth of 28~kHz. In this frequency range, only zero-order symmetric, $S_0$, anti-symmetric, $A_0$, and shear $SH_0$ modes can propagate in the plate. The emitter is not able to produce shear waves, and will almost only excite the first anti-symetric mode, as it generates anti-symetric displacement. The presence of discrete steps may induce small conversion of anti-symetric modes into other propagation modes, but we can simply consider this as a part of the losses occuring in the system. A laser vibrometer (Polytec sensor head OFV534, controller OFV2500) is used to measure the acoustic velocity of the flexural waves. The vibrometer is only sensitive to vertical-displacement velocity. At low frequency ($kh\ll 1$) the symetric mode is almost an in-plane wave \cite{royer2000}, therefore we measure preferentially the anti-symmetric mode. The laser probe is scanned over the flat surface of the plate with a $1~mm-$step grid to map the complete spatio-temporal velocity field. 

\begin{figure}[]
\begin{center}
\includegraphics[]{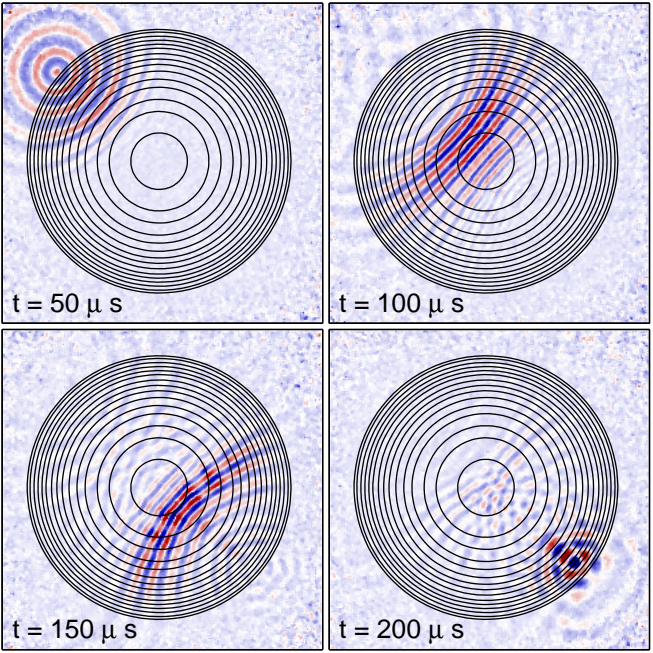}
\end{center}
\mbox{}\vspace{-1cm}
\caption{Experiment: Velocity field distribution measured at four different times: $t$ = 50, 100, 150, and 200~$\mu$s. The source is at the top left and emits a Gaussian pulse centered at $f_0$ = 60~kHz. The time origin is set at the beginning of emission. Red lines: positive amplitudes; blue lines: negative amplitudes. Color scale is identical in all four snapshots. (Multimedia view)}
\label{fig_manip60kHz}
\end{figure}

Figure \ref{fig_manip60kHz} (multimedia view) shows snapshots of the measured field at four different times. Red and blue lines represent successive positive and negative wavefronts. As time progresses, wavefront concavity changes from positive around the source to negative around the image point. After time $t= 200~\mu s$, the initial pulse refocuses at a point diametrically opposed to the source, as predicted theoretically. The  thickness gradient realizes an almost perfect Maxwell fish-eye lens without the need for reflecting or clamped boundaries. Similar results are obtained for a source located at different points on the outer disk. The translation of thickness variation into index gradient is further illustrated by the wavefront separation which decreases toward the center of the acoustic lens where thickness as well as phase velocity are the smallest. Halfway of pulse propagation, wavefronts are flat. This confirms that a half lens would convert a point source located at the edge of the lens into a plane wave and \emph{vice versa}. One might notice that focusing occurs slightly inside the lens. This is attributed to the discrepancy between ideal and actual thickness profiles (Fig.\ref{fig_design}). We confirmed this hypothesis numerically (see supplementary material \cite{SuppMat}). Because the  actual lens is thinner than computed, the phase velocity is reduced along the diagonal trajectory and focusing is reached earlier within the plate.

\begin{figure}[]
\begin{center}
\includegraphics[]{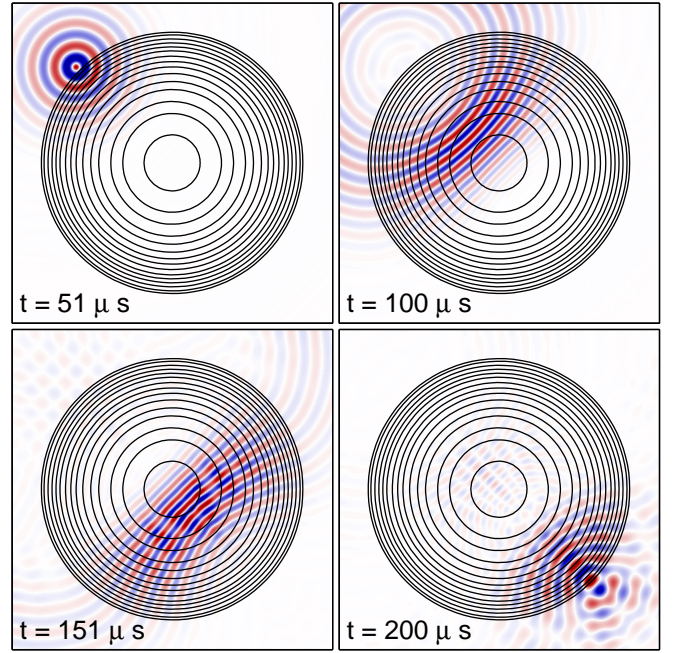}
\end{center}
\mbox{}\vspace{-1cm}
\caption{Numerical simulations using 3D-FDTD algorithm: Velocity field distribution measured at four different times. Same initial conditions as in Fig.~\ref{fig_manip60kHz}. The origin of time is set at the beginning of emission. Red lines, positive amplitudes; blue lines, negative amplitudes. Color scale is identical in all four snapshots.}
\label{fig_simu60kHz}
\end{figure}

Exact numerical simulations of flexural waves propagation in a system similar to the experimental setup have been performed based on three dimensional finite difference time-domain method (3D-FDTD) using Simsonic3D software \cite{Virieux1986, bossy2004}. We apply stress-free boundary conditions on the upper and lower surfaces of the plate and consider that the plate stands in vacuum. We use highly absorbing layers as lateral boundary conditions, which eliminate reflected waves. Figure \ref{fig_simu60kHz} shows four snapshots of the velocity field at about the same time steps as in Fig.~\ref{fig_manip60kHz}). Refocusing is observed around time $t=200~\mu s$, close to the experiment. The presence of highly absorbing layers confirms that focusing observed experimentally is not the result of residual reflections in the square cavity, but the result of wavefront shaping by the ultrasonic lens.

\begin{figure}[]
\begin{center}
\includegraphics[]{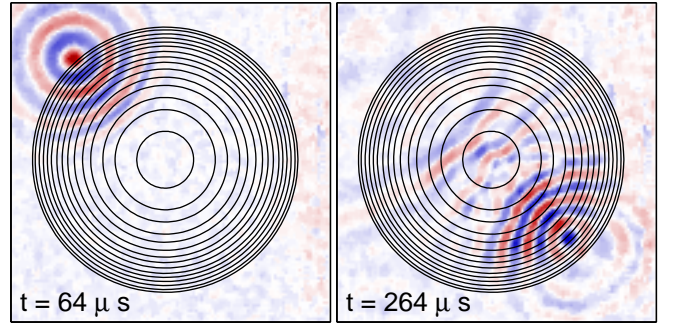}
\end{center}
\mbox{}\vspace{-1cm}
\caption{Experiment: Velocity field distribution measured at two different times. The source is located at the top left corner and emits a gaussian pulse centered at $f_0$=30~kHz. Red lines, positive amplitude; blue lines, negative amplitude. Color scale is identical in the two snapshots. }
\label{fig_manip30kHz}
\end{figure}

Focusing is also demonstrated at lower central frequency $f_0$=30~kHz in Fig.~\ref{fig_manip30kHz}. Importantly, the lensing effect is broadband. The limit of applicability of geometric acoustics gives the lower bound. For the range of plate thicknesses which we use, the wavelength becomes comparable to the lens size for carrier frequency $f_0$ around 1 kHz. At the other end of the spectrum, the upper limit is given by the validity of the quadratic approximation for the dispersion relation Eq.~\ref{eq_dispersion}. This relation is valid for $kh \leq 1$, which means Eq.~\ref{eq_dispersion} breaks down above $f_0$=300~kHz in our case. Acoustic lensing by our Maxwell fish-eye is therefore expected to be efficient over two frequency decades, which is a broader frequency range than its electromagnetic counterpart \cite{ulf2011}. This is attributed to the unique dispersive nature of flexural waves in thin plates.

\begin{figure}[]
\begin{center}
\includegraphics[]{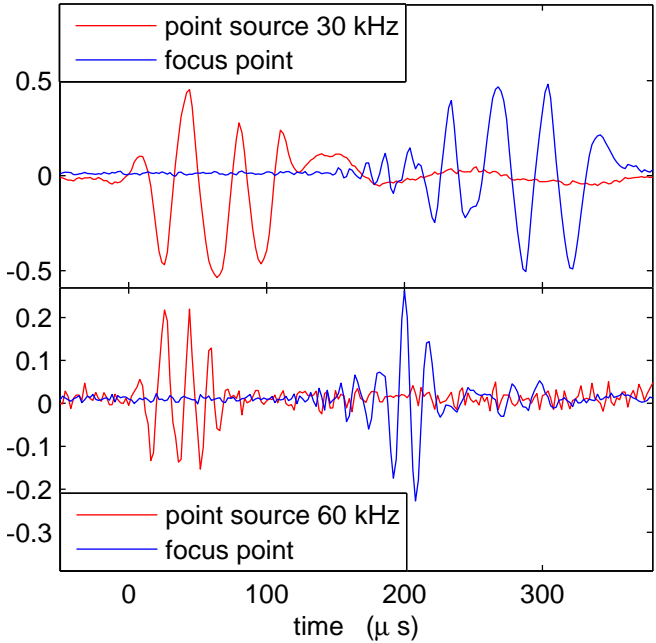}
\end{center}
\mbox{}\vspace{-1cm}
\caption{Signal measured at the source position and at the focus point. The pulse travels from the source to the focusing point in (a) 200~$\mu$s at 30~kHz, and (b) 150~$\mu$s at 60~kHz. This time delay is measured at the center of the gaussian pulse.}
\label{fig_time_flight}
\end{figure}

The impact of dispersion is also revealed in the dynamics. Refocusing time is found to be dependent on the carrier frequency. Figure \ref{fig_time_flight} illustrates this point. At $30$~kHz, it takes around 200~$\mu$s for the pulse to converge on the other side of the lens, while this delay is reduced by 25$\%$ to 150~$\mu$s when carrier frequency is doubled. This difference can be estimated by recalling that the ultrasonic rays follow projections on the plane of meridians of the equivalent sphere. As a result, the plate diagonal and the external perimeter have actually the same acoustic length as they correspond to shortest paths (geodesics). It is therefore equivalent and therefore easier to compute the time of flight over the external perimeter where the plate thickness is constant. The impulsion propagates at the group velocity $V_g$ which is simply twice the phase velocity in the particular case of a quadratic dispersion relation. Thus we obtain a refocusing time of 203~$\mu$s at 30~kHz, and 144~$\mu$s at 60~kHz, in good agreement with experimental observations.
A direct outcome of this frequency dependence of the refocusing time is the possibility to separate temporally different spectral components at a single focusing point.

In conclusion, we have studied numerically and experimentally the dynamics of the Maxwell fish-eye lens transposed to flexural waves and based on a thin duraluminium plate with varying thickness. Such a Maxwell's fish-eye makes a paradigm for transformation based lenses. This lensing effect has proven to be very resilient, as it persists despite a limited accuracy of the experimental device fabrication. We stress that analysis of time focusing of Lamb waves can also help to solve the controversy on whether or not Maxwell's fish-eye makes a perfect lens \cite{ulf2009,ulf2011,merlin2011}, but this is beyond the scope of the present work as it requires a plate with a circular stress-free boundary. Moreover, other thickness variations can be envisaged \cite{dehesa2014}, what makes for instance Eaton lenses for flexural waves possible. We note that such lensing effects might be observed with pillar based metamaterial plates \cite{Rupin2014} with a variation of pillars's height like in Eq. \ref{eq_thickness}. Finally, an extension of our work to Rayleigh waves for pillars on semi-infinite elastic substrates \cite{Achaoui2013} with a variation of height like in Eq. \ref{eq_thickness} would open interesting routes
towards lensing of surface elastic waves in seismic metamaterials based on
analogies with transformation optics.

M.D. acknowledges Ph.D. funding from the Direction G\'en\'erale de l'Armement (DGA). P.S. is thankful to the Agence Nationale de la Recherche support under grant ANR PLATON (No. 12-BS09-003-01), the LABEX WIFI (Laboratory of Excellence within the French Program Investments for the Future) under reference ANR-10-IDEX-0001-02 PSL* and the Groupement de Recherche 3219 MesoImage. 
S.G. wishes to thank ERC for funding through ANAMORPHISM grant.

%\bibliography{Fisheye_Biblio}

\end{document}

\title{Experiments on Maxwell's Fish-eye dynamics in elastic plates \\ Supplementary Materials}
\maketitle

\renewcommand{\thefigure}{S\arabic{figure}}

\setcounter{figure}{0}

\section*{Additional simulation}

\begin{figure}[h!]
\begin{center}
\includegraphics[scale=1]{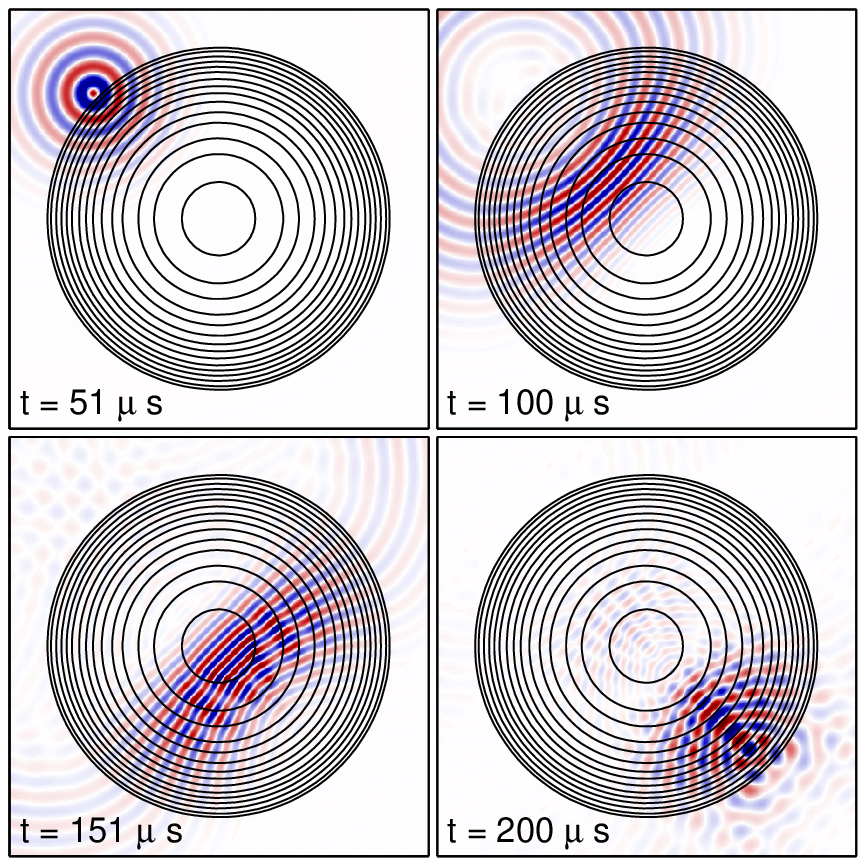}
\includegraphics[scale=1]{manip60kHz.eps}
\end{center}
\caption{Left: Simulation of a thinner plate, whose thickness starts from 400 $\mu$m in the center to 1.9 mm on the edges. The focusing occurs slightly inside the lens, just as in the experiment. Right: The corresponding experiment is displayed aside to stress the similarity of the results.}
\label{fig_simu}
\end{figure}

In the accompanying letter, we have demonstrated pulse refocusing by a platonic Maxwell fisheye lens. However, we found that focusing occurs within the lens and not at its edge as predicted. We attributed this effect to the difference between the fabricted lens profile and the theoretical profile. In order to confirm this hypothesis, we performed numerical simulations with a plate profile similar to the actual profile used in the experiment and thinner than the theoretical one. The same thickness variation has ben used but with a global shift of 100 $\mu$m. Figure \ref{fig_simu} compares the new simulation and the experiment. In both cases the focal image spot is located inside the lens, which confirms the hypothesis. We note that flexural wave focussing by Maxwell's fish-eye is a robust effect and variation from theoretical lens profile does not hinder the effect but simply shifts the focal point.  

\section*{Adiabatic thickness variation}

As stated in the introduction of the accompanying letter, the thickness variation of the plate needs to be slow for Kirchoff-Love equation to be valid. The generalized equation for plates of variable thickness $h$ is: 

\begin{multline}\label{eq_Generalized-Kirchoff}
\rho h \frac{\partial^2 W}{\partial t^2}
+ D \Delta^2 W + \Delta D \Delta W 
+ 2 \frac{\partial D}{\partial x} \frac{\partial }{\partial x} \Delta W	 
+ 2 \frac{\partial D}{\partial y} \frac{\partial }{\partial y} \Delta W\
- (1-\nu)\left(\frac{\partial^2 D}{\partial x^2}\frac{\partial^2 W}{\partial y^2}-2\frac{\partial^2 D}{\partial x \partial y}\frac{\partial^2 W}{\partial x \partial y}
+\frac{\partial^2 D}{\partial y^2}\frac{\partial^2 W}{\partial x^2}\right) =0.
\end{multline}
where $D=E h^3/12(1-\nu^2)$ is the flexural rigidity, with $E$ the Young's modulus and $\nu$, the Poisson's ratio of the plate, respectively. Besides from that, $\rho$ is the plate density and $W$ its vertical displacement.

We can argue on the basis of dimensional analysis that the thickness variation is slow enough to consider Kirchoff-Love plate equation
\begin{equation}\label{Love}
\rho h \frac{\partial^2 W}{\partial t^2}
+ D \Delta^2 W =0.
\end{equation}
Indeed, in that case $D \Delta^2 W\gg\Delta D \Delta W$
%The variation of $D$ gives birth to additional terms of the form $\Delta D \Delta W$, or $\nabla D \nabla^3 W$, which are to be compared with $D \Delta^{2} W$.
since the vertical displacement $W$ varies on a scale of the flexural wave wavelength $\lambda$, whereas flexural rigidity varies over the dimension of the lens $R_{max}$. Indeed, if we evaluate the order of magnitude of the ratio of these terms,
we find that:
\begin{equation}\label{eq_ratio_ordre2}
	\frac{|\Delta D \Delta W|}{|D \Delta^2 W|} \sim 	\left(\frac{\lambda}{R_{max}}\right)^2,
\end{equation}
which is small compared to $1$ since in our experiments we
typically have a ratio $\lambda / R_{max} \sim $ 0.1.

Similarly $D \Delta^2 W$
dominates all the remaining terms of Eq. \ref{eq_Generalized-Kirchoff}. For instance
one has
\begin{equation}\label{eq_ratio_ordre1}
	\frac{|\nabla D \nabla^3 W|}{|D \Delta^2 W|} \sim \frac{\lambda}{R_{max}},
\end{equation}
which is again a small quantity as long as the geometric acoustics approximation is valid (although Eq. \ref{eq_ratio_ordre2} requires an even less restrictive
condition than Eq. \ref{eq_ratio_ordre1} to be negligible).
If we were to refine the approximate
Eq. \ref{Love}, the term
$\nabla D \nabla^3 W$ should be added in order
to get a higher-order approximation of Eq. \ref{eq_Generalized-Kirchoff}, the
term $\Delta D \Delta W(\ll\nabla D \nabla^3 W)$ would still remain negligible.
Such an asymptotic analysis of the problem is nonetheless beyond the scope of
the present work, and it would not alter the physical discussion.